# Enhanced injection efficiency and light output in bottom tunnel-junction light-emitting diodes using UID GaN spacers

Shyam Bharadwaj[1,2], Jeffrey Miller[1], Kevin Lee[1], Joshua Lederman[1], Marcin Siekacz[2], Huili (Grace) Xing[1,3,4], Debdeep Jena[1,3,4], Czesław Skierbiszewski[2], and Henryk Turski[2]

[1]Department of Electrical and Computer Engineering, Cornell University, Ithaca, NY 14853 USA,

[2]Institute of High Pressure Physics, Polish Academy of Sciences, Sokołowska 29/37, PL-01-142 Warsaw, Poland

[3]Department of Material Science and Engineering, Cornell University, Ithaca, NY 14853 USA,

[4]Kavli Institute for Nanoscale Science, Cornell University, Ithaca, NY 14853 USA

*Abstract:*

*Recently, the use of bottom-TJ geometry in LEDs, which achieves N-polar-like alignment of polarization fields in conventional metal-polar orientations, has enabled enhancements in LED performance due to improved injection efficiency. Here, we elucidate the root causes behind the enhanced injection efficiency by employing mature laser diode structures with optimized heterojunction GaN/In$_{0.17}$Ga$_{0.83}$N/GaN TJs and UID GaN spacers to separate the optical mode from the heavily doped absorbing p-cladding regions. In such laser structures, polarization offsets at the electron blocking layer, spacer, and quantum barrier interfaces play discernable roles in carrier transport. By comparing a top-TJ structure to a bottom-TJ structure, and correlating features in the electroluminescence, capacitance-voltage, and current-voltage characteristics to unique signatures of the N- and Ga-polar polarization heterointerfaces in energy band diagram simulations, we identify that improved hole injection at low currents, and improved electron blocking at high currents, leads to higher injection efficiency and higher output power for the bottom-TJ device throughout 5 orders of current density (0.015 – 1000 A/cm$^2$). Moreover, even with the addition of a UID GaN spacer, differential resistances are state-of-the-art, below 7x10$^{-4}$ $\Omega$cm$^2$. These results highlight the virtues of the bottom-TJ geometry for use in high-efficiency laser diodes.*

Indium Gallium Nitride (InGaN) based light-emitting diodes (LEDs) and laser diodes (LDs) have garnered much attention for applications in visible lighting, displays, and light fidelity (Li-Fi) communications[1,2,3,4,5,6]. With direct bandgaps spanning the visible wavelength range, the InGaN material system has enabled high efficiency visible and white lighting, with external quantum efficiencies (EQEs) reaching 90% for blue and violet emitters[7,8]. However, the hole injection layers in such devices, typically consisting of GaN or AlGaN, suffer from large activation energies >200 meV for the Mg acceptor, increasing series resistance and limiting the wall-plug efficiency (WPE). The resulting low p-type conductivities are especially problematic in standard p-up geometry LDs – in such devices, tensile strain limits the permissible Al mole fraction in the claddings, and thus limits the refractive index contrast between the cladding and active region. To then adequately confine the optical mode, thick low-Al AlGaN claddings are required, which contribute large series resistances to the devices[9,10]. In addition to generating resistive electrical losses, these thick heavily doped p-claddings induce optical absorption losses, which stem from a high level of absorbing unionized Mg acceptor-bound holes present in these layers[11]. The high Mg concentration coupled with significant optical mode overlap result in the p-cladding layer being the largest contributor to optical loss in LDs[12,13]. Furthermore, metal contacts covering much of the resistive p-layer are required for current spreading. Absorption of the optical mode by the large-area metal p-contact may contribute to optical losses and therefore increase the lasing threshold current even further[14].

Several methods have been investigated to mitigate optical and electrical losses in LDs. The use of thick undoped waveguide or spacer regions on either side of the active region to limit leakage of the optical mode into heavily doped layers (especially p-type) has been shown to reduce optical absorption losses[12,15]. Additionally, many groups have incorporated tunnel junctions (TJs) into LEDs and LDs in order to reduce electrical losses from p-type contacts and layers[9,10,16,17,18]. Using a TJ, consisting of a heavily doped p-n homojunction or heterojunction, can offer significant improvements over standard p-type layers. In TJ devices, n$^+$GaN serves as the anode contact layer in place of p$^+$GaN, lowering the resistivity by ~2 orders, allowing for enhanced current spreading with reduced metal contact area. In LDs, this allows for placement of the contact away from the laser ridge; in such geometry, refractive index contrast between the top cladding and the waveguide is increased (as air serves as a cladding with n=1), allowing for enhanced optical confinement, and for thinning down the resistive p-GaN or p-AlGaN cladding[9,10]. More recently, a study by Turski et al. on the alignment of polarization and p-n junction fields in TJ LEDs has shown that utilizing bottom-TJ geometry, with the TJ grown beneath a p-down LED (achieving N-polar-like alignment of polarization fields, but grown in the metal-polar direction) results in higher injection efficiencies and greater light output than using the standard top-TJ geometry[19].

This study expands on the prior bottom-TJ LED result by comparing injection efficiency (defined here as the ratio of total recombination current in the QW active region to total recombination current in the structure) between improved bottom-TJ and top-TJ laser diode epi-structures. Compared to the structures in reference 1, the structures in this study are more suitable for high efficiency laser diodes due to the presence of an unintentionally doped (UID) GaN spacer separating the heavily doped p-layers from the active region, and a low-resistance heterojunction $GaN/In_{0.17}Ga_{0.83}N/GaN$ TJ instead of a GaN homojunction TJ. Detailed measurements of electroluminescence (EL, spanning nearly 5 orders of injected current density: between 0.015 A/cm$^2$ and 1000 A/cm$^2$), current-voltage (IV) characteristics, and capacitance-voltage (CV) characteristics elicit the root causes for higher injection efficiency in the bottom-TJ structure over the entire range of investigated current densities. The results, when correlated with SiLENSe simulations of the devices, show that improved hole injection is responsible for the increased efficiency prior to device turn-on, beyond which improved electron blocking becomes the main cause for the increased efficiency. The state-of-the-art differential resistances measured in these TJ devices despite presence of a UID spacer (5.4x10$^{-4}$ Ωcm$^2$ for the top-TJ and 6.4x10$^{-4}$ Ωcm$^2$ for the bottom-TJ device at 1000 A/cm$^2$) coupled with the stronger electroluminescence from the bottom-TJ device suggest that for metal-polar TJ laser diodes with UID spacer, the optimal geometry is p-down, rather than the commonly used p-up geometry.

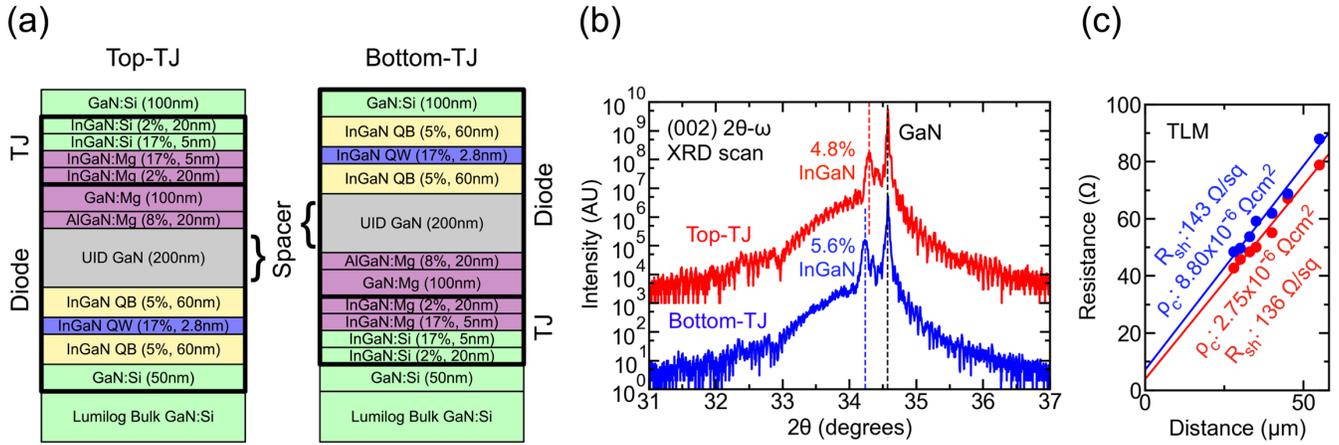

Fig. 1: (a) Schematic structures for the two samples in this study, denoting the pn-diode region (used in SiLENSe simulations) and the TJ region. (b) XRD data for the two samples, showing good agreement in layer thicknesses and compositions between the structures. The slight shift to the left for the peak at ~34.25 degrees in the bottom-TJ structure indicates ~0.8% higher In composition in the InGaN QB regions compared to the top-TJ structure. (c) TLM measurements for the top contact for both structures after fabrication into diodes. Resulting contact and sheet resistances are similar for both devices, and acceptably low for contacts to n-GaN ($R_{sh} < 150\ \Omega$, $\rho_c < 10^{-5}\ \Omega cm^2$).

To study the effects of polarization field orientation on LD injection efficiency and light output, two structures were grown by plasma-assisted molecular beam epitaxy (MBE) on Ga-polar bulk n-GaN substrates (schematic structures shown in Fig. 1(a)). The top-TJ structure begins with growth of an n-GaN nucleation layer followed by growth of the active region: a 2.8 nm $In_{0.17}Ga_{0.83}N$ quantum well sandwiched by 60 nm $In_{0.05}Ga_{0.95}N$ quantum barriers (QBs). Then, a 200 nm UID GaN spacer layer is grown to separate subsequent p-layers from the active region. Next, a 20 nm $Al_{0.08}Ga_{0.92}N$ electron blocking layer (EBL) is grown, followed by the p-GaN cladding, a GaN/$In_{0.17}Ga_{0.83}N$/GaN TJ, and finally an $n^+$GaN contact layer. The bottom-TJ structure consists of the same layers but grown in the opposite order, with the exception of the nucleation and top contact layers, which are in the same positions as in the top-TJ. Each corresponding layer of the top-TJ and bottom-TJ structures is grown at the same substrate temperature and using the same Al/Ga/In/N flux to make fair comparisons. Compositions and thicknesses were confirmed by X-ray diffraction (XRD – Fig. 1(b)), confirming consistency in layer structures between the two samples. The slight difference in 2-theta angle for the InGaN QB peak at 34.25 degrees can be attributed to 0.8% unintentionally higher In composition in the bottom-TJ device's InGaN QB layers. From SiLENSe simulations, this slightly higher In composition is not expected to affect device performance significantly – in fact, it is expected to *reduce* injection efficiency slightly, and therefore does not preclude meaningful comparisons of device performance. Further, the heterostructure TJ used here, with an $In_{0.17}Ga_{0.83}N$ interlayer, reduces depletion width more in the top-TJ geometry than in bottom-TJ geometry in which an AlN interlayer would theoretically reduce depletion width most effectively[20].

After structural characterization, the samples were co-processed into LEDs in order to perform electrical and optical characterization. Processing devices in this manner (rather than into laser stripes) allows for simple extraction of EL, CV, and IV characteristics and comparison between different devices without the need to decouple factors such as differences in mirror losses. LED mesas were defined through inductively-coupled plasma reactive ion etching (ICP-RIE). For both samples, etch depth extends into the substrate, such that TJ cross sectional area is uniform between the samples. After mesa isolation, top and back contacts consisting of Ti/Al = 25/100 nm were deposited through electron beam evaporation. The top contact was annealed at 550 °C for 1 minute, resulting in contact resistances in the $10^{-6}$ $\Omega cm^2$ range. Devices with areas ranging from 20x20 $\mu m^2$ to 300x300 $\mu m^2$ were realized, with all of the device measurements in this study performed on 80x80 $\mu m^2$ devices. Measured contact and sheet resistances by the transfer length method (TLM) are similar between the two devices, with slightly lower values for the top-TJ device (Fig. 1(c)). Thus, it is unlikely that the benefits seen in the bottom-TJ structure derive from differences in device fabrication; they are due to the orientation of polarization fields alone. In addition to the measurements, simulations of the output power, CV, and IV characteristics were performed using SiLENSe. The simulated structures exclude the TJ region (indicated in Fig 1(a)) – since the additional series resistance due to inclusion of a TJ is low in comparison to other series resistances present in the structures (especially that of the UID GaN spacer), this method provides a good approximation of the true device. The results of the simulations, and their high quality of fit to the measurements, suggest that the limiting characteristics of the devices are determined by the structure of the diode region, and that the bottom-TJ device layer quality is not significantly affected by growth on top of a heavily doped TJ.

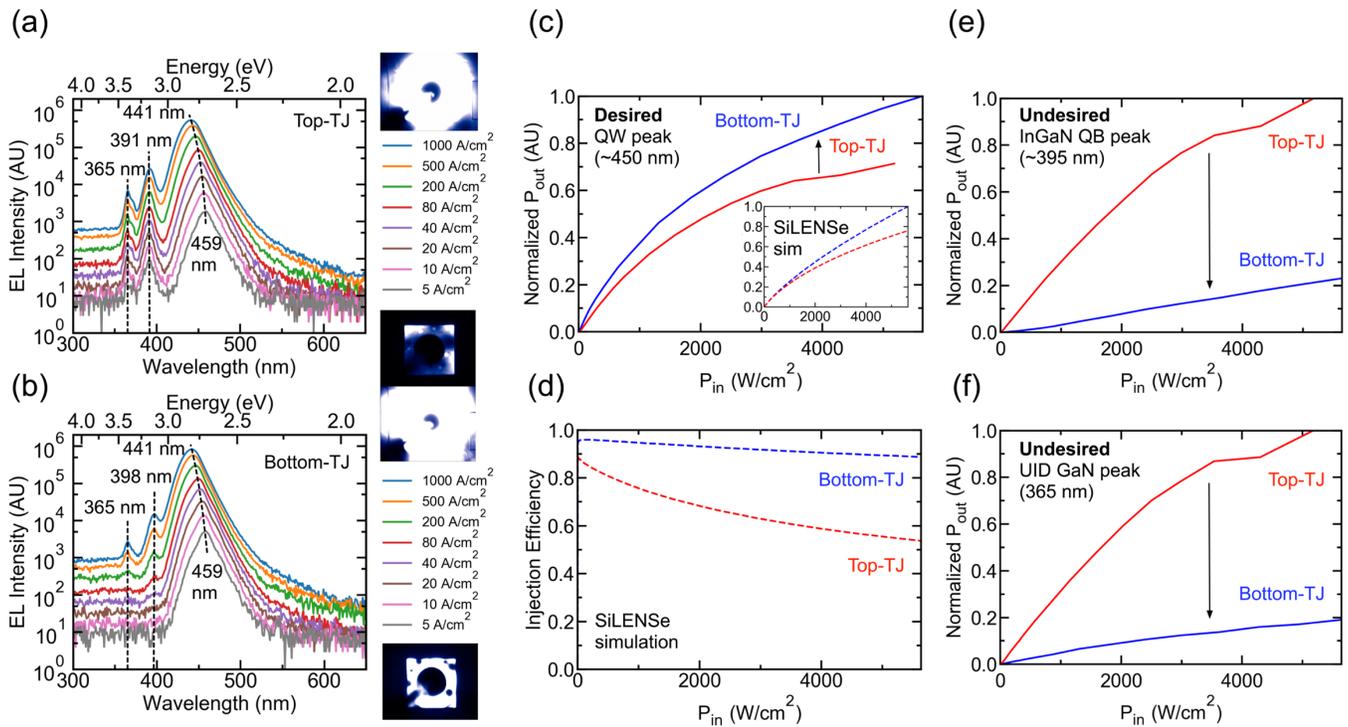

Fig. 2: (a), (b) EL spectra from 80x80 μm² devices between 5 A/cm² and 1000 A/cm² (top-TJ in (a), bottom-TJ in (b)), with false color images of the devices at the low and high current densities to the right of the plots. The main InGaN QW peak occurs at 459-441 nm (blueshifting with increased current density), and is stronger for the bottom-TJ device. Parasitic peaks are present at ~395 nm and 365 nm, coming from the InGaN QB and UID GaN spacer regions, respectively. (c) Integrated EL intensity vs input power for the main peak for both devices, showing good agreement with the SiLENSe simulation in the inset. (d) Simulated injection efficiency for the two devices. (e), (f) Integrated EL intensity vs input power for the two parasitic peaks, showing ~5x lower parasitic emission in the bottom-TJ device compared to the top-TJ device.

After processing the epi-structures into devices, EL measurements were performed at current densities between 0.015 A/cm² and 1000 A/cm² on 80x80 μm² devices to compare emission intensities and wavelengths. A subset of the collected spectra is shown in Figs. 2(a) and (b). It can be seen that the main emission peak from the active region occurs at 459 nm at low injection current (5 A/cm²), blueshifting to 441 nm at 1000 A/cm² for both samples. The identical main peak location confirms identical QW thickness and composition between the two samples. Parasitic emission peaks are seen at ~395 nm and 365 nm, coming from recombination in the InGaN QB region and the UID GaN spacer respectively. With increasing current density, first the parasitic peak at ~395 nm appears, and later the peak at 365 nm, with intensity for the ~395 nm peak always stronger by ~5x. This behavior suggests that electron overshoot is responsible for the parasitic emission, increasing in severity with increased current injection, with more electrons present in the regions where the conduction band edge is lower in energy. The parasitic peaks are always present in the top-TJ sample even below 5 A/cm$^2$, but emerge only at high current densities in the bottom-TJ sample (the InGaN QB peak appears at ~40 A/cm$^2$, and then the GaN spacer peak at ~200 A/cm$^2$), suggesting that the orientation of polarization fields in bottom-TJ geometry improves electron blocking, as shown later in Fig. 4(e). The slight difference in peak wavelength for the ~395 nm emission peak between the two samples (391 nm for top-TJ and 398 nm for bottom-TJ) is due to the ~0.8% compositional difference in the InGaN QBs, as seen from XRD. Compared to previous results on other blue bottom-TJ LEDs which showed no parasitic peaks[19], the higher current densities investigated here cause some degree of carrier leakage even in the bottom-TJ structure, albeit much smaller than in the top-TJ.

A comparison of the integrated EL intensities of the three observed peaks for the two samples as output power density ($P_{out}$ (A.U.)) vs input power density ($P_{in}$ = J*V) is shown in Figs. 2(c)-(f). For the two parasitic emission peaks (Figs. 2(e), (f)), the intensity at a fixed input power is stronger for the top-TJ structure by >5x throughout the range of input powers. In contrast, for the desired QW peak (Fig. 2(c)), the intensity is stronger in the bottom-TJ structure by ~1.3x. SiLENSe simulations (inset of Fig. 2(c)) of the output power predict this trend for intensity of the QW peak: the intensity is always higher for the bottom-TJ structure, reaching ~1.35x that of the top-TJ at an input power of 5000 W/cm$^2$. The disparity in $P_{out}$ between the two structures increases with increasing $P_{in}$, reflecting the similar trend seen in the simulated injection efficiency (Fig. 2(d)): the injection efficiency is lower for the top-TJ structure, with the difference increasing with increased $P_{in}$. Internal quantum efficiency (IQE, defined as the ratio of radiative recombination current in the QW to total recombination current in the QW) is similar for both structures, leaving injection efficiency as the root cause for differences in device performance. The increase in relative intensity for the main QW seen in the bottom-TJ sample with respect to the top-TJ sample (showing good agreement with simulation), coupled with decrease in intensity for parasitic emission peaks, is a signature of improved electron blocking in the bottom-TJ device when the devices are turned on.

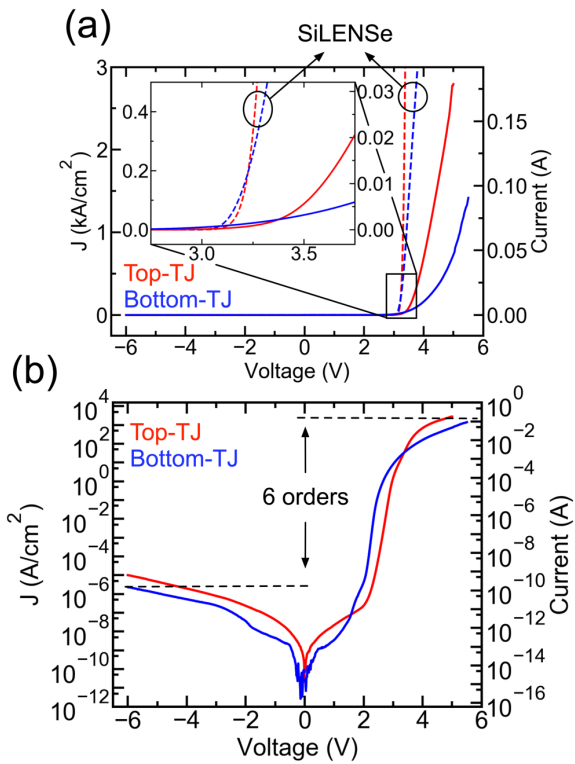

Fig. 3: (a), (b) Linear and log-scale IV characteristics, respectively, for the devices in this study, with SiLENSe simulations of the IV characteristics shown in (a). Voltage and current for the bottom-TJ device are flipped in sign for easy comparison. The bottom-TJ device shows higher current than the top-TJ prior to turn-on, and lower current after, consistent with simulations. Differential resistances at 1000 A/cm$^2$ are 5.4x10$^{-4}$ Ωcm$^2$ for the top-TJ and 6.4x10$^{-4}$ Ωcm$^2$ for the bottom-TJ device. The devices show similar levels of leakage current, with ~6 orders of rectification.

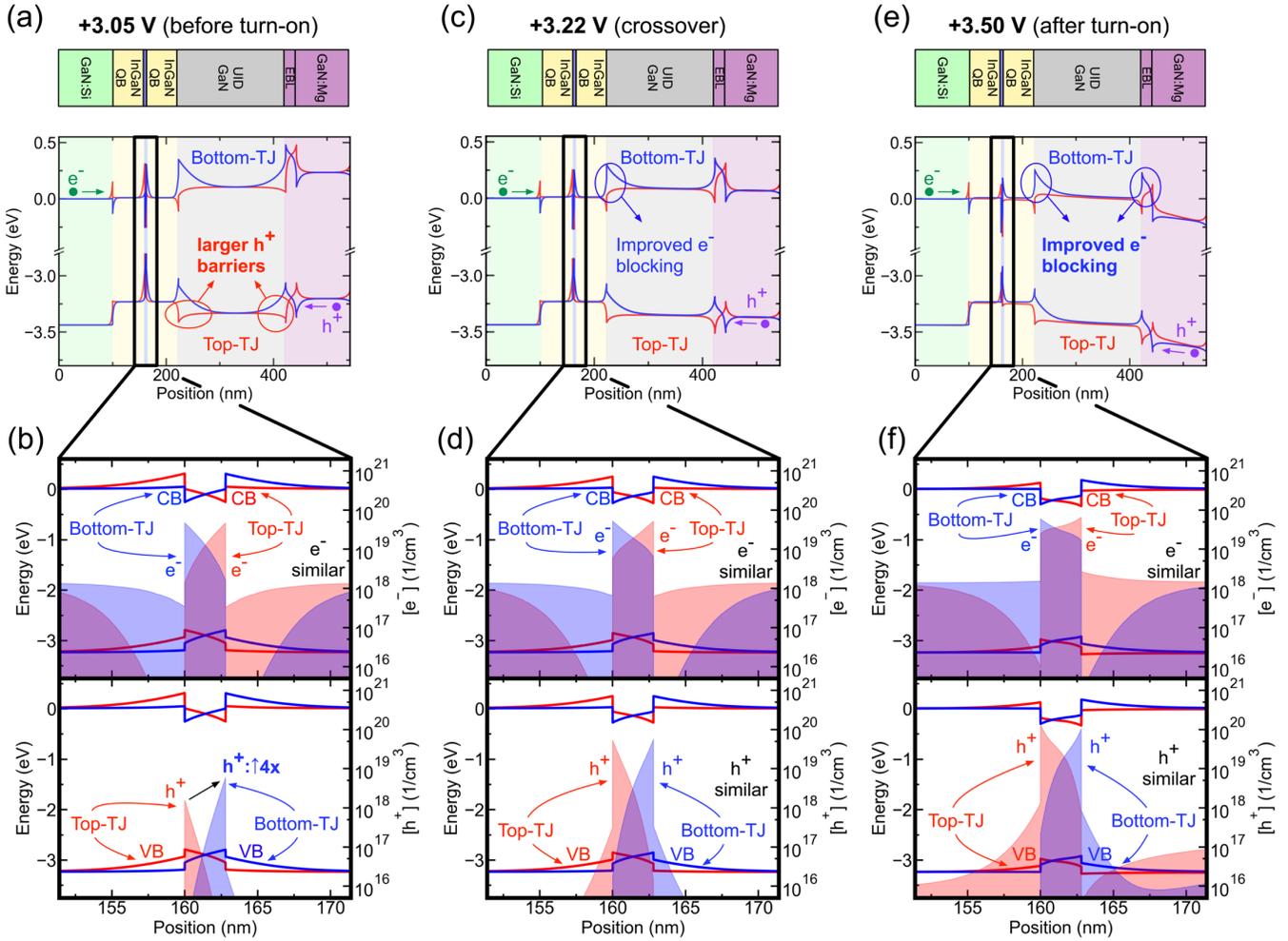

Fig. 4: (a), (b) Band diagram and carrier concentrations in the InGaN QW, respectively, at 3.05 V forward bias (before device turn-on). Larger barriers to hole injection are seen for the top-TJ device, resulting in 4x lower hole concentrations in the QW and lower current compared to the bottom-TJ device. (c), (d) Band diagram and carrier concentrations in the InGaN QW, respectively, at 3.22 V forward bias (at the crossover point). Current density is ~182 A/cm$^2$ for both devices. The barriers to hole injection seen in (a) for the top-TJ device are sufficiently screened, leading to similar hole concentrations in the QW as in the bottom-TJ device. (e), (f) Band diagram and carrier concentrations in the InGaN QW, respectively, at 3.50 V forward bias (after device turn-on). At this bias, the bottom-TJ device now has significantly larger barriers to electron overflow than the top-TJ device.

The differences in carrier injection and the injection efficiency between the bottom-TJ and top-TJ devices can be further understood by investigating the measured and simulated IV and CV characteristics, and how they relate to the simulated energy band diagrams. The IV characteristics are shown in Fig. 3, energy band diagrams in Fig. 4., and CV in Fig. 5. The bottom-TJ IV data is plotted with voltage and current flipped in sign for easy comparison. Qualitatively, from the IV data, it is seen that the measured current in the bottom-TJ structure is higher than in the top-TJ until ~3.4 V, after which the current is lower for the bottom-TJ device (Fig. 3(a), measured data). Simulated IV from SiLENSe (also shown in Fig. 3(a)) predicts this "crossover" behavior: higher current for the bottom-TJ device until ~3.2 V, after which the top-TJ shows higher current. Complementing the IV data, CV measurements reveal differences in device behavior prior to turn-on. An increased zero-bias capacitance and a shoulder in the CV characteristic is observed for the bottom-TJ. This shoulder is absent in the top-TJ CV near turn on (Figs. 5(f), (e)). The calculated energy band diagrams and carrier concentrations from SiLENSe (Figs. 5(a) – (d)) provide an explanation for this difference in behavior.

The CV characteristics were extracted from impedance measurements over the 80 kHz to 700 kHz range at DC biases out to 3 V forward bias. In this measurement window, the impedance is almost entirely reactive (-90 < phase < -80 deg.); hence, a small AC signal ($V_{AC}$ = 30 mV) superimposed onto the applied DC bias electrostatically modulates electron and hole populations within the structures, and any current flow (with phase = 0 deg.) contributes negligibly to the measured impedance. In this way, differences in the magnitude and spatial extent of the electron and hole populations in the bottom-TJ and top-TJ structures, owing to the flip in polarization, arise as variations in extracted capacitance. There is a slight dispersion in frequency owing to the small, but finite, current flow in the LDs before turn-on. As the capacitive impedance scales as 1/f, the contribution from the parallel conductance to the total (measured) conductance is comparatively smaller with increasing frequency. Hence, the electrostatic approximation is improved with higher frequencies, with the measured phase pushed closer to -90 degrees at higher frequencies over the range. As more current begins to flow beyond ~2.5 V, the phase sharply increases towards 0 degrees in both devices, and the electrostatic approximation is no longer valid. In this regime the measured impedance sharply decreases owing to the rapid increase in current flow from both devices; hence the extracted capacitance is no longer accurate.

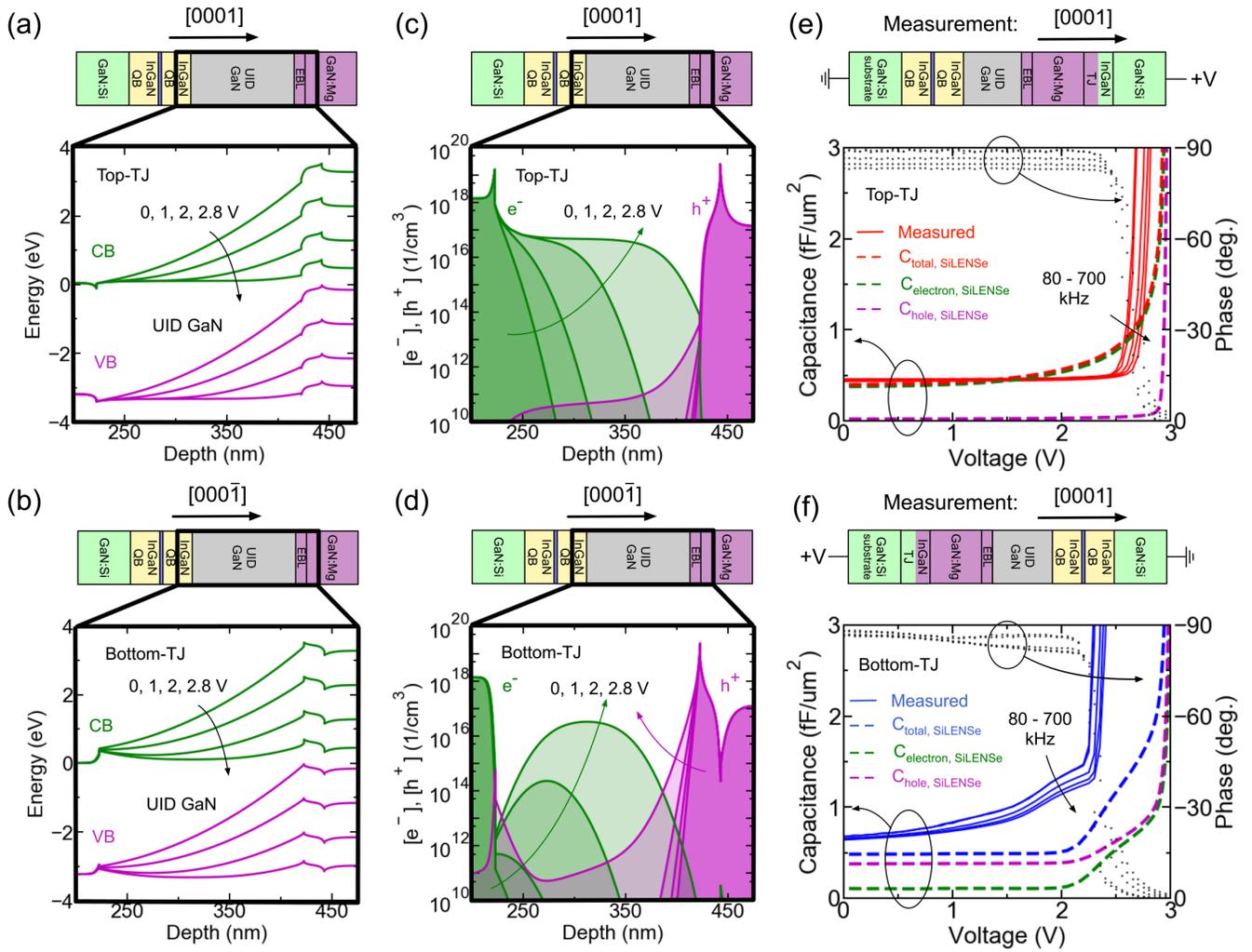

Fig. 5: (a), (b) Band diagrams for the top-TJ and bottom-TJ structures, respectively, from 0 – 2.8 V forward bias, zoomed into the UID GaN spacer region. (c), (d) Carrier concentrations for the top-TJ and bottom-TJ structures, respectively, at the biases that band diagrams are shown for in (a) and (b). For the top-TJ, the bands gradually flatten out, and electrons move uniformly into the UID spacer, with very little hole modulation. In contrast, the bands bow rather than flatten for the bottom-TJ, allowing more hole modulation than for the top-TJ. (e), (f) Measured and simulated CV characteristics for the top- and bottom-TJ devices, respectively. Measurements are performed at 80, 100, 300, 500, and 700kHz. Until ~2.5 V, the phase angle remains acceptably close to -90 degrees. The top-TJ CV is relatively constant, due to the uniform extension of electrons with increasing bias. The bottom-TJ CV is more complex, with non-uniform carrier extension and contributions from both electrons and holes causing a shoulder to appear.

To help quantify and explain the observed trends in IV and CV, selected data at 3.05 V (before the current crossover), 3.22 V (at the crossover), and 3.50 V (after the crossover) simulated with SiLENSe is shown in Table 1. Looking first at the voltages before the crossover, the simulations reveal that current flow is dominated by radiative current stemming from recombination in the QW ($J_{rad, QW}$), and to a lesser degree by nonradiative recombination current in the QW ($J_{nrad, QW}$). At these biases, $J_{nrad, QW}$ constitutes a similar fraction of total current for the two devices, and electron injection into the QW is also similar. The main difference between the two structures is in the hole injection into the QW as indicated in Figs. 4(a) and (b). Holes see larger barriers in the top-TJ geometry (Fig. 4(a)) from the AlGaN EBL/UID GaN spacer interface and the UID GaN spacer/InGaN quantum barrier interface because of the orientation of polarization fields. On top of this, additional holes are present in the bottom-TJ QW due to the InGaN QB/InGaN QW polarization field, while instead, electrons are present in the top-TJ at the same location. With overall hole concentration much lower than electron concentration at these biases (Fig. 4(b)), the difference in hole injection becomes the critical reason that current flow is higher in the bottom-TJ at biases below ~3.2 V.

| Sample | Bias (V) | $J_{total}$ (A/cm$^2$) | $J_{rad, QW}$ (A/cm$^2$, % of $J_{total}$) | $J_{nrad, QW}$ (A/cm$^2$, % of $J_{total}$) | $J_{anode}$ (A/cm$^2$, % of $J_{total}$) – carrier overflow | WPE (%) |
|---|---|---|---|---|---|---|
| Top-TJ | 3.05 | 1.14 | **0.604 (53%)** | 0.471 (41%) | 0.059 (5%) | 48% |
|  | 3.22 | 183 | **84.3 (46%)** | 67.2 (37%) | 31.5 (17%) | 39% |
|  | 3.50 | 11,300 | 1,100 (10%) | 2080 (18%) | **8130 (72%)** | 7.6% |
| Bottom-TJ | 3.05 | 4.17 | **2.64 (63%)** | 1.474 (36%) | 0.054 (1%) | 57% |
|  | 3.22 | 182 | **97.2 (53%)** | 78.5 (43%) | 5.3 (3%) | 46% |
|  | 3.50 | 1,430 | 528 (37%) | **769.3 (53%)** | 133 (9%) | 29% |

Table 1: Selected data from SiLENSe showing the contributions to total current flow for the top- and bottom-TJ devices at 3.05 V, 3.22 V, and 3.50 V – the three biases for which band diagrams are shown in Fig. 4(a), (c), and (e), respectively. All material parameters were kept constant for the two devices. Although the values listed may change with the material parameters, the observed trends remain. The largest contributor to total current is shown in bold for each device at each bias. Prior to device turn-on (3.05 V) as well as at the crossover (3.22 V), radiative and nonradiative current in the QW are the largest contributors for both devices, with slightly higher radiative current in the bottom-TJ device. Notably, at the crossover point, with voltage and $J_{total}$ the same for both devices, the bottom-TJ device still shows enhanced $J_{rad, QW}$ and suppressed $J_{anode}$. After turn-on (3.50 V), $J_{rad, QW}$ remains a significant contributor for the bottom-TJ device, but is much lower for the top-TJ device. At this bias, $J_{anode}$ stemming from electron overflow is the dominant contributor to total current for the top-TJ device. The final column shows calculated wall-plug efficiency at each bias, assuming that all emission is at 2.75 eV (450 nm). WPE for the bottom-TJ device is always higher than that of the top-TJ.

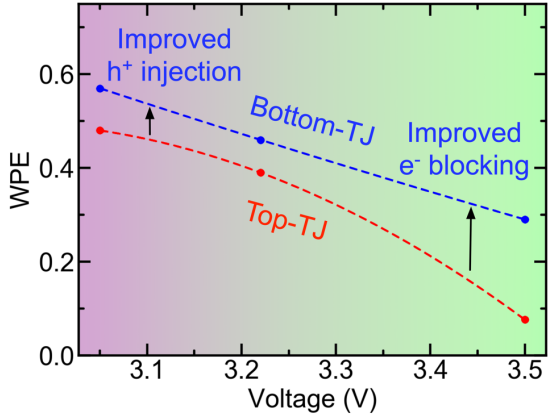

Fig. 6: Calculated WPE versus forward bias voltage, using the points from the final column in Table 1. Different regimes are depicted in the plot with shading to denote the different causes for higher WPE in the bottom-TJ structure. At lower biases, improved hole injection is responsible for the increased WPE, while at higher biases, improved electron blocking leads to higher WPE.

Electrostatic simulations of CV support this conclusion: the capacitance of the top-TJ before turn-on is well described by the electron population alone (Fig. 5(e)). On the other hand, hole modulation is crucial in reproducing the measured bottom-TJ CV characteristic (Fig. 5(f)). Prior to device turn-on, the CV simulations show that the charge modulation almost entirely occurs within the UID GaN spacer, and therefore simulated band diagrams and carrier populations zoomed into this region are shown in Figs. 5(a)-(d). For small increments in DC bias ($\Delta V$, 100 mV step size), we compute the change in electron density, $\Delta n(x)$, and hole density, $\Delta p(x)$, within the spacer region from SiLENSe. The total capacitance, then, is the sum of individual electron and hole $\frac{\Delta Q}{\Delta V}$ components:

$$C_{total} = \frac{\Delta Q}{\Delta V}\bigg]_{V=V_{dc}} = C_{electron} + C_{hole} = \frac{q \int \Delta n(x) dx}{\Delta V} + \frac{q \int \Delta p(x) dx}{\Delta V},$$

where q is the elementary charge. For the top-TJ, the orientation of the polarization field at the UID GaN spacer/InGaN QB interface allows the conduction band in the spacer to flatten out just before turn on, and a roughly uniform electron population extends from the interface into the spacer. Simultaneously, the polarization field pulls the valence band down at the AlGaN EBL/UID GaN spacer interface, limiting hole buildup (Figs. 5(a) and (c)). The low hole population results in $C_{electron}$ dominating over $C_{hole}$ for the top-TJ, while the growing extension of electrons into the spacer results in an exponential-like $C_{total}$. For the bottom-TJ, the polarization field at the AlGaN EBL/UID GaN spacer interface allows for larger hole concentrations in the spacer (Fig. 5(d)). Hence, $C_{hole}$ is a large component of the capacitance at low bias. The polarization fields at the two interfaces cause bowing of the UID spacer bands (shown in Fig. 5(b)), contrasting the flat bands in the top-TJ, resulting in non-uniform carrier populations appearing in the spacer. In particular, the simulations suggest electrons accumulate in the center of the spacer around 2 V. While holes dominate the capacitance at low voltages, this accumulation of electrons at a later bias causes $C_{total}$ to exhibit the shoulder, as seen in the measurement.

For biases at the crossover point (Figs. 4(c) and (d)) and higher (Figs. 4(e) and (f)), the hole barriers in the top-TJ are sufficiently screened by the applied field. In addition, the lack of polarization-induced holes at the InGaN QB/InGaN QW interface in the top-TJ device becomes less consequential due to the large number of holes generated by the applied field. These effects result in similar hole concentrations in the QW for both devices (slightly higher for top-TJ at 3.50 V). Interestingly, at the crossover point, even with identical voltage and current across both devices (identical $P_{in}$), the bottom-TJ device exhibits higher $J_{rad, QW}$, as expected from Fig. 2(c). Additionally, differences in electron blocking can be seen in Fig. 4(c) at the UID GaN spacer/InGaN QB interface for the two devices. Already, electron overflow is becoming a problem in the top-TJ device, with recombination current at the anode ($J_{anode}$) at 3.22 V reaching 17% of total current ($J_{total}$) as opposed to only 3% for the bottom-TJ device (Table 1). This problem becomes much worse in the top-TJ device at +3.50 V: $J_{anode}$ accounts for >70% of the total current, while $J_{rad, QW}$ accounts for <10%. The large anode current is unsurprising given the electron buildup in the UID spacer even at the lower biases. In contrast, while simulations show intrinsically higher on-resistance for the bottom-TJ beyond the crossover, $J_{anode}$ accounts for <10% of the total current flow, while $J_{rad, QW}$ accounts for 37%. Now, the polarization fields generated at both the UID GaN spacer/InGaN QB and AlGaN EBL/UID GaN spacer interfaces contribute to this higher on-resistance and improved electron blocking (Fig. 4(e)). Calculated WPE for the bottom-TJ device is always higher than that of the top-TJ even at the higher biases at which on-resistance is higher, as seen in the last column of Table 1. The higher measured on-resistance for the bottom-TJ after the devices turn on is seemingly in agreement with the simulation, though it is not likely that we have achieved series resistances low enough to reveal the intrinsically higher bottom-TJ diode resistance at these biases, and therefore there is room to improve the bottom-TJ on-current up to the levels of what is seen for the top-TJ. However, at the lower biases in which the intrinsic diode resistance dominates, the agreement between the simulation and measurement is meaningful, showing the intrinsically superior hole injection in the bottom-TJ compared to top-TJ. A summary of the different mechanisms leading to higher WPE for the bottom-TJ device in the corresponding bias regimes in which they dominate is shown in Fig. 6.

In conclusion, we show that using bottom-TJ geometry in a LD structure with a UID GaN spacer results in an enhancement in light output power of ~30% in comparison to a standard top-TJ device. It is notable as well that the differential resistances (5.4 x $10^{-4}$ $\Omega$ cm$^2$ for the top-TJ and 6.4 x $10^{-4}$ $\Omega$ cm$^2$ for the bottom-TJ device at 1000 A/cm$^2$) and current levels seen in these devices are on par with the best results for TJ LEDs found in literature to date despite the presence of a UID spacer[21,22,23]. Through SiLENSe simulations and measurements of light output power, CV, and IV characteristics, we identify different mechanisms for the enhanced light output in different injection current (or equivalently, different $P_{in}$) regimes. In the low current regime, enhanced hole injection in the bottom-TJ device is responsible for the improved device performance. At the crossover point and after the devices turn on, differences in performance stem mainly from improvements in electron blocking. The higher differential resistance for the bottom-TJ compared to the top-TJ suggests there is room for improvement in terms of electrical conductivity. Towards this end, taking advantage of polarization-induced doping in the form of compositional grading in the cladding layers may reduce overall series resistance. Furthermore, the use of an AlN interlayer rather than In$_{0.17}$Ga$_{0.83}$N theoretically reduces the depletion width in the TJ even more for the bottom-TJ device, so it is possible that further optimization of the TJ region may improve device performance[20]. Additionally, this work has not utilized the inherent advantage offered by the bottom-TJ geometry: that of the TJ cross sectional area. Since device isolation only requires etching through the active region, the TJ region can remain unetched for the bottom-TJ, while it must be etched for the top-TJ. The much larger achievable TJ cross sectional area in bottom-TJ geometry will result in a far lower tunneling resistance. These methods provide a straightforward path to significantly improving bottom-TJ device performance further, and will allow for enhanced laser performance in the future.

Acknowledgements:

This work was supported partially by the Polish National Centre for Research and Development (Grant No. LIDER/29/0185/L-7/15/NCBR/2016) during the visit of the lead author at Institute of High Pressure Physics PAS and the Foundation for Polish Science co-financed by the European Union under the European Regional Development Fund (Grant. No. POIR.04.04.00-00-5D5B/18-00). The work at Cornell University was supported in part by the following National Science Foundation (NSF) grants: NSF DMREF Award No. 1534303 monitored by Dr. J. Schluter, NSF Award No. 1710298 monitored by Dr. T. Paskova, NSF CCMR MRSEC Award No. 1719875, and NSF RAISE TAQs Award No. 1839196 monitored by Dr. D. Dagenais. This work made use of the shared facilities which are supported through the NSF National Nanotechnology Coordinated Infrastructure (Grant No. ECCS-1542081), NSF MRSEC program (DMR-1719875), and MRI DMR-1338010.